\documentclass[prc,preprint,showpacs,showkeys,nofootinbib]{revtex4}%
\usepackage{graphicx}
\usepackage{bm}
\usepackage{epsfig}
\usepackage{amsmath}
\usepackage{amsfonts}
\usepackage{amssymb}%
\begin{document}
\title{Is the tetraneutron a bound dineutron-dineutron molecule?}
\author{C.A. Bertulani and V. Zelevinsky}
\email{bertulani@nscl.msu.edu, zelevinsky@nscl.msu.edu}
\affiliation{National Superconducting Cyclotron Laboratory, and Department of Physics and
Astronomy, Michigan State University, East Lansing, MI\ 48824-1321, USA}
\date{\today}

\begin{abstract}
In light of a new experiment which claims a positive identification, we
discuss the possible existence of the tetraneutron. We explore a model based
on a dineutron-dineutron molecule. We show that this model is not able to
explain the tetraneutron as a bound state, in agreement with other theoretical
models already discussed in the literature.

\end{abstract}
\maketitle

\draft

\address{National Superconducting Cyclotron Laboratory, and
Department of Physics and Astronomy, Michigan State University, E.
Lansing, MI 48824-1321}

\narrowtext

In a recent experiment \cite{Ma02}, the existence of bound neutron clusters
was studied by fragmentation of intermediate energy (30-50 MeV/nucleon)
$^{14}$Be nuclei. In particular, the fragmentation channel $^{10}%
\mathrm{Be}+\ ^{4}\mathrm{n}$ was observed and the $^{4}\mathrm{n}$ system was
tentatively described as a bound tetraneutron system.

The possible existence of the tetraneutron has been discussed theoretically by
numerous authors already in the 1960's (e.g., refs.
\cite{Ar62,Gol63,Loh63,Vla64,Hip64,Ben64,Dan65,Tan65}). The experimental
search at that time \cite{Sch63,Bri64,Cie65,Coh65}, including the experiments
on double pion charge exchange reactions with $^{4}$He \cite{Dav64,Gil65},
gave negative results. The general conclusion, based on experiment and theory,
was that the tetraneutron cannot be bound. More recent experiments
\cite{Ung84,Bel88,Ale88,Gor89} and theoretical calculations
\cite{Bev80,Bad85,Gorr89,Var95,Sof97} give support to this assertion, except
perhaps ref. \cite{Gut89} where a hyperspherical function method was used and
led the authors to the conclusion that the tetraneutron may exist as a
resonance in the four-body continuum at energy of about 1-3 MeV. These
theoretical models have recently been reconsidered in ref. \cite{Tim02} where
it was concluded that a too strong four-nucleon force is needed to bind the
tetraneutron and that this force would unreasonably bind $^{4}$He by about 100 MeV.

It thus seems very unlikely that the tetraneutron can be explained within any
standard theoretical model. In this brief report we consider the possibility
that the tetraneutron can be described as a composite dineutron-dineutron
molecular system. To our knowledge this is a hitherto unexplored model for the
possible binding of neutron matter. In this model, the dineutron is considered
to be slightly bound due to a polarization mechanism induced by the presence
of the other dineutron. Although the dineutron system is not bound (it has a
virtual singlet state at $E=66$ keV), there is some theoretical
\cite{Mi73,Pud96,Sme97} and experimental \cite{Sim99} evidence that it might
become bound in the presence of another nuclear system.

Another reason to believe that neutrons clusterize, or form correlated bound
pairs inside a nucleus, is the existence of the Borromean nuclear systems,
e.g. $^{6}$He and $^{11}$Li, or nultineutron configuration systems in light
nuclei, e.g. $^{8}$He. The Borromean systems \cite{Zh93} are thee-body systems
in which the particles are not bound in pairs, but the three-body system is
bound. For example, although $^{10}$Li and the two-neutron systems are not
bound, $^{11}$Li is bound. It is the presence of $^{10}$Li which induces the
neutron-neutron binding correlation, as first suggested by Migdal \cite{Mi73}.
The authors of ref. \cite{Zhuk94} also suggest that two of the three most
likely ground state configurations in $^{8}$He are compatible with an
$\alpha+^{4}$n cluster system.

\begin{figure}[ptb]
\begin{center}
\includegraphics[
height=1.9086in,
width=3.0493in
]{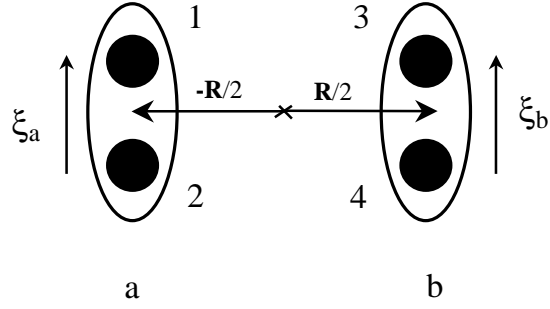}
\end{center}
\caption{Tetraneutron as a dineutron-dineutron molecule.}%
\label{dd}%
\end{figure}

The model for the tetraneutron is based on two clusters of dineutron molecules
as displayed in Fig. \ref{dd}. A similar model was used in the framework of
the generator coordinate method for the scattering problem in ref.
\cite{Gir73}. The molecular models, similar to those used in quantum
chemistry, were successfully applied to the chain of Be isotopes \cite{Kan02}.
The dineutron wave function can be written as%
\begin{equation}
\varphi_{12}=\sqrt{2}\mathcal{A}_{12}\left\{  N_{0}\exp\left[  -\frac{\xi
_{a}^{2}}{2b^{2}}\right]  \chi_{1}^{\left(  +\right)  }\chi_{2}^{\left(
-\right)  }\right\}  , \label{gausswf}%
\end{equation}
where $\xi_{a}=2\left\vert \mathbf{r}_{1}+\mathbf{R}/2\right\vert $ is the
(relative) intrinsic coordinate and $\mathbf{r}_{1}$\ is the position of the
neutron 1 with respect to the center of mass of the tetraneutron. The relative
motion spatial wave function is taken as a Gaussian with $N_{0}=\left(
b^{3}\pi^{3/2}\right)  ^{-1/2}$, and $b$ is the oscillator parameter. The
dineutron is assumed to be in a spin-0 state, described by the spinors
$\chi_{1}^{\left(  +\right)  }$ (spin-up) and $\chi_{2}^{\left(  -\right)  }$
(spin-down), respectively. The wave function is antisymmetrized by the
operator $\mathcal{A}_{12}=\frac{1}{2}\left(  1-P_{12}\right)  $, where
$P_{12}$\ exchanges neutrons 1 and 2. An analogous wave function,
$\varphi_{34}$, is written for the second dineutron molecule. Since the
spatial part of the dineutron internal wave functions is even under inversion,
the operator $\mathcal{A}_{12}$ only acts over the spin variables and we can
write Eq. (\ref{gausswf}) as%
\begin{align}
\varphi_{12}  &  =N_{0}\exp\left[  -\frac{\xi_{a}^{2}}{2b^{2}}\right]
\frac{1}{\sqrt{2}}\left\{  \chi_{1}^{\left(  +\right)  }\chi_{2}^{\left(
-\right)  }-\chi_{1}^{\left(  -\right)  }\chi_{2}^{\left(  +\right)  }\right\}
\nonumber\\
&  =N_{0}\exp\left[  -\frac{\xi_{a}^{2}}{2b^{2}}\right]  \chi_{12}^{\left(
0\right)  }. \label{nnwf2}%
\end{align}
Accordingly, the wave function for the second dineutron is%
\begin{equation}
\varphi_{34}=N_{0}\exp\left[  -\frac{\xi_{a}^{2}}{2b^{2}}\right]  \chi
_{34}^{\left(  0\right)  }. \label{nnwf3}%
\end{equation}

The total wave function of the tetraneutron is%
\begin{equation}
\Psi=\sqrt{6}\mathcal{A}_{12,34}\left[  \varphi_{12}\varphi_{34}\right]
\Phi(\mathbf{R}), \label{tetrawf}%
\end{equation}
where the operator $\mathcal{A}_{12,34}=\frac{1}{6}\left(  1-P_{13}%
-P_{14}-P_{23}-P_{24}+P_{13}P_{24}+P_{14}P_{23}\right)  $ implements the
antisymmetrization between the dineutrons. The factors $\sqrt{2}$ in Eq.
(\ref{gausswf}) and $\sqrt{6}$\ in Eq. (\ref{tetrawf}) account for the proper
normalization. The function $\Phi\left(  \mathbf{R}\right)  $\ is the
dineutron-dineutron molecular wave function. 

Using Eqs. (1-4), we obtain (here we drop the upper index (0) in the singlet
spinors)%
\begin{equation}
\Psi=\sqrt{6}N_{0}^{2}\left\{  f\left(  \xi_{a},\xi_{b}\right)  \chi_{12}%
\chi_{34}-g\left(  \mathbf{\mbox{\boldmath$\xi$}}_{a}%
,\mathbf{\mbox{\boldmath$\xi$}}_{b},\mathbf{R}\right)  \chi_{14}\chi
_{32}-h\left(  \mathbf{\mbox{\boldmath$\xi$}}_{a}%
,\mathbf{\mbox{\boldmath$\xi$}}_{b},\mathbf{R}\right)  \chi_{13}\chi
_{24}\right\}  \Phi(\mathbf{R}), \label{Psi2}%
\end{equation}
where%
\begin{equation}
f\left(  \xi_{a},\xi_{b}\right)  =\frac{1}{2}\exp\left[  -\frac{\xi_{a}%
^{2}+\xi_{b}^{2}}{2b^{2}}\right]  , \label{nf1}%
\end{equation}%
\begin{equation}
g\left(  \mathbf{\mbox{\boldmath$\xi$}}_{a},\mathbf{\mbox{\boldmath$\xi$}}%
_{b},\mathbf{R}\right)  =\frac{2}{3}f\left(  \xi_{a},\xi_{b}\right)
\ \exp\left[  -\frac{R^{2}}{b^{2}}\right]  \exp\left[  \frac
{\mathbf{\mbox{\boldmath$\xi$}}_{a}.\mathbf{\mbox{\boldmath$\xi$}}_{b}}%
{2b^{2}}\right]  , \label{ng1}%
\end{equation}
and
\begin{equation}
h\left(  \mathbf{\mbox{\boldmath$\xi$}}_{a},\mathbf{\mbox{\boldmath$\xi$}}%
_{b},\mathbf{R}\right)  =\frac{2}{3}f\left(  \xi_{a},\xi_{b}\right)
\ \exp\left[  -\frac{R^{2}}{b^{2}}\right]  \exp\left[  -\frac
{\mathbf{\mbox{\boldmath$\xi$}}_{a}.\mathbf{\mbox{\boldmath$\xi$}}_{b}}%
{2b^{2}}\right]  . \label{nh1}%
\end{equation}

The total Hamiltonian for the tetraneutron system is%
\begin{equation}
H=-\frac{\hbar^{2}}{2m_{N}}\sum_{i=1}^{4}\Delta_{i}+V\ . \label{Hamilt}%
\end{equation}
The neutron-neutron potential was taken as a two-body Volkov potential
\cite{Vo65},%
\begin{equation}
V={\displaystyle\sum\limits_{i\leq j}^{4}} \left(  1-M+MP_{ij}^{x}\right)
V_{ij}\ ,\ \ \ \ \ \ \ \ V_{ij}\left(  r\right)  =V_{\alpha}\exp\left(
-r^{2}/\alpha^{2}\right)  +V_{\beta}\exp\left(  -r^{2}/\beta^{2}\right)  ,
\end{equation}
where $P_{ij}^{x}$\ \ is the Majorana exchange operator and the parameters
$V_{\alpha}$, $V_{\beta},$ $\alpha$ and $\beta$ are chosen to reproduce the
scattering length and effective range in low energy nucleon-nucleon
collisions, as well as the binding energy of $^{4}\mathrm{He}$\ \cite{Vo65}.

\begin{figure}[t]
\begin{center}
\includegraphics[
height=3in, width=3in ]{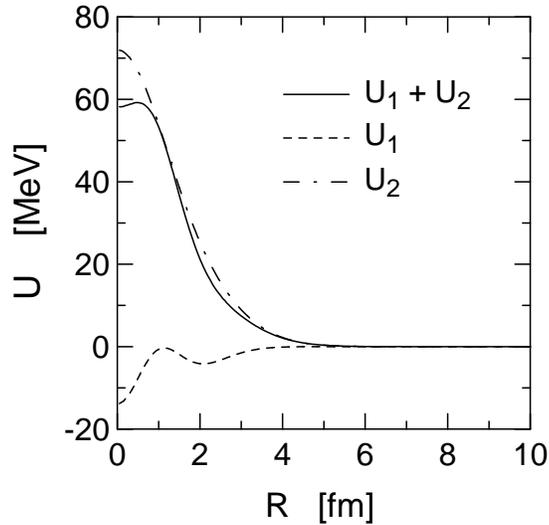}
\end{center}
\caption{The effective potentials $U_{1}(R)$ and $U_{2}(R)$ entering Eq.
(\ref{effective}). The solid curve is the sum of the two potentials.}%
\label{g2}%
\end{figure}

The total Hamiltonian $H$ can be presented as%
\begin{equation}
H=T_{a}+T_{b}+T_{R}+V, \label{H1}%
\end{equation}
with%
\begin{equation}
T_{a,b}=-\frac{\hbar^{2}}{4m_{N}}\mathbf{\mbox{\boldmath$\triangle$}}%
_{\mathbf{\mbox{\boldmath$\xi$}}_{a,b}},\ \ \ \ \ \ T_{R}=-\frac{\hbar^{2}%
}{2m_{N}}\mathbf{\mbox{\boldmath$\triangle$}}_{\mathbf{R}}. \label{H2}%
\end{equation}
Since in our model the mean positions of the clusters are taken to be at
$\mathbf{R}/2$ and $-\mathbf{R}/2$, respectively, and since the position of
each nucleon is symmetric with respect to the origin of each cluster, there is
no admixture of spurious motion of the center of mass.

\begin{figure}[ptb]
\begin{center}
\includegraphics[
height=3.8in, width=3.2in ]{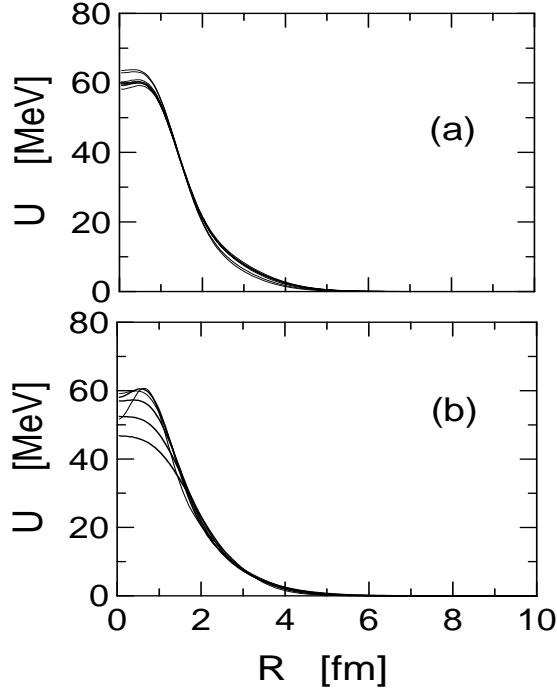}
\end{center}
\caption{ (a) The total potential $U_{1}(R)+U_{2}(R)$ entering Eq.
(\ref{effective}) for 8 different parameterizations of the Volkov potential.
The oscillator parameter $b=1.5$ fm was used. (b) The same as in (a), but
using the Volkov-1 interaction and varying $b$ from 1.2 fm to 2.0 fm.}%
\label{g3}%
\end{figure}

We use Eq. (\ref{Psi2}) as a starting point of a variational procedure. To
obtain an effective Schr\"{o}dinger equation for $\Phi\left(  \mathbf{R}%
\right)  $, we multiply the full 4-body Schr\"{o}dinger equation $H\Psi=E\Psi$
by $\Psi^{\dagger}/\Phi\left(  \mathbf{R}\right)  $ and integrate over
$\mbox{\boldmath$\xi$}_{1}$ and $\mbox{\boldmath$\xi$}_{2}$.\ \ We will assume
the variational function $\Phi\left(  \mathbf{R}\right)  $ to be independent
of angles (i.e., an \textit{s-wave} state),%
\begin{equation}
\Phi\left(  \mathbf{R}\right)  =\frac{1}{\sqrt{4\pi}}\frac{u(R)}{R}\ .
\label{swave}%
\end{equation}
so that the operator $T_{R}$ in eq. (\ref{H2}) becomes%
\begin{equation}
T_{R}=-\frac{\hbar^{2}}{2m_{N}}\frac{d^{2}}{dR^{2}}. \label{H3}%
\end{equation}
The integrals over $\mbox{\boldmath$\xi$}_{1}$ and $\mbox{\boldmath$\xi$}_{2}$
can be performed analytically. One obtains an effective Schr\"{o}dinger
equation for $u(R)$\ in the form%
\begin{equation}
-\frac{\hbar^{2}}{2m_{N}}u^{\prime\prime}(R)+RV_{3}(R)u^{\prime}(R)+\left\{
V_{1}(R)+V_{2}(R)-E\right\}  u(R)=0 \label{almost1}%
\end{equation}
where%
\begin{align}
V_{1}(R)  &  =\frac{\hbar^{2}}{m_{N}b^{2}}\left(  \frac{R^{2}}{b^{2}}-\frac
{7}{4}\right)  \exp\left(  -R^{2}/b^{2}\right)  \ ,\label{v1cor}\\
V_{2}(R)  &  =2\left[  M-1-M\frac{1-\exp\left(  -2R^{2}/b^{2}\right)  }%
{1-\exp\left(  -R^{2}/b^{2}\right)  }\right]  \left[  f\left(  \alpha\right)
+f\left(  \beta\right)  \right]  \ ,\label{v2cor}\\
f\left(  \alpha\right)   &  =\frac{V_{\alpha}\alpha^{3}}{\left(  \alpha
^{2}+b^{2}\right)  ^{3/2}}\left\{  1+\exp\left[  -\frac{R^{2}}{\alpha
^{2}+b^{2}}\right]  \right\}  \ , \label{v3cor}%
\end{align}
and%
\begin{equation}
V_{3}(R)=-\frac{2\hbar^{2}}{m_{N}b^{2}}\exp\left(  -R^{2}/b^{2}\right)  .
\label{v4cor}%
\end{equation}
Since we are interested only in the relative energy between the clusters, we
extract from the total Hamiltonian the internal kinetic energy of each
cluster: $T^{int}=3\hbar^{2}/\left(  4mb^{2}\right)  $\ and $V^{int}%
=V_{\alpha}\left(  1+b^{2}/\alpha^{2}\right)  ^{-3/2}+V_{\beta}\left(
1+b^{2}/\beta^{2}\right)  ^{-3/2}$ that do not depend on $R$.

Eq. (\ref{almost1}) can be put in a conventional form of a Schr\"{o}dinger
equation:%
\begin{equation}
-\frac{\hbar^{2}}{2m_{N}}v^{\prime\prime}(R)+\left\{  U_{1}(R)+U_{2}%
(R)-E\right\}  v(R)=0, \label{effective}%
\end{equation}
where%
\begin{equation}
v(R)=u(R)\exp\left[  \frac{1}{2}\int_{0}^{R}f(R^{\prime})dR^{\prime}\right]
,\ \ \ \ \ \ \ f(R)=-\frac{\hbar^{2}}{2m_{N}} R V_{3}(R) \ . \label{v5cor}%
\end{equation}
The effective potentials in eq. \ref{effective} are given by
\begin{equation}
U_{2}(R)=V_{2}(R), \label{v6cor}%
\end{equation}
and%
\begin{align}
U_{1}(R)  &  =V_{1}(R)-\frac{RV_{3}^{\prime}(R)}{2} -{\frac{V_{3}(R)}{2}}
+\frac{m_{N}}{2 \hbar^{2}}\left[  RV_{3}(R)\right]  ^{2}\nonumber\\
&  =-\frac{\hbar^{2}}{m_{N}b^{2}}\left[  \left(  \frac{3}{4}+\frac{R^{2}%
}{b^{2}}\right)  \exp\left(  -R^{2}/b^{2}\right)  -\frac{2R^{2}}{b^{2}}%
\exp\left(  -2R^{2}/b^{2}\right)  \right]  . \label{v7cor}%
\end{align}

We use for the oscillator parameter $b=1.5$ fm. A set of 8 parameters of the
Volkov potential were taken from Ref. \cite{Vo65}. In Fig. \ref{g2} we show
the results for the Volkov V1 force. We notice that the potential $U_{2}%
(R)$\ is repulsive and the potential $U_{1}(R)$\ is attractive. Their sum is
dominated by the repulsive part.

In Fig. \ref{g3}(a) we show the total potential $U_{1}(R)+U_{2}(R)$\ for the 8
sets of parameters in the Volkov potential, as taken from Ref. \cite{Vo65}.
One sees that none of the parameter sets leads to a potential with an
attractive pocket, which could be a sign for the existence of a bound state.
But even in that case the pocket would have to be deep enough to allow for the
appearance of the bound state.

Finally, we have varied the oscillator parameter from 1.2 fm to 2 fm for each
set of parameters of the Volkov interaction. No pocket appeared in the
effective potential $U_{1}(R)+U_{2}(R)$ within this range of variation. Figure
\ref{g3}(b) shows this for the specific case of the Volkov-1 interaction.

In summary, we have explored a model of the tetraneutron as a
dineutron-dineutron molecule. Using a variational calculation we have found an
effective Schr\"{o}dinger equation for the relative motion of the dineutrons,
after a proper account for the Pauli exclusion principle. An effective
potential for the relative motion of the dineutron molecules was obtained. We
showed that this potential does not have a pocket and thus the tetraneutron is
very unlikely to be bound as a dineutron-dineutron molecule, although more
complex variational approaches still can be explored. For example, one might
consider a different spatial wavefunction for the dineutron system. We have
used Gaussian for convenience. But the asymptotic form of the wavefunction
might not be correct. An Yukawa form might be more appropriate to achieve a
lower energy for the system. Another improvement might be the use of more
realistic interactions, other than the Volkov interaction. Interactions
including tensor parts might lead to some modifications of our results.
Finally, although less probable as an improvement, one might relax the
assumption of singlet states for the dineutron allowing for the triplet
configurations in the calculation.

Our study is complementary to other approaches and reinforces the commonly
accepted idea that a tetraneutron is not a possible outcome of a theoretical
calculation starting with underlying two-body nucleon-nucleon interactions. If
the tetraneutron is bound, most probably it will be due to a special four-body
attraction in $T=2$ states or an exceptional fine tuning of the
nucleon-nucleon interaction which does not seem to fit within present nuclear
models. It might however be useful to recall an old phenomenological argument
\cite{Vla64} against the stability of the tetraneutron. Adding a pair of
neutrons to a nucleus one usually increases the separation energy of the
proton. If this rule holds, a simple comparison of the particle-stable tritium
and unstable $^{5}$H immediately leads to the conclusion that $M(^{4}%
\mathrm{n})>4M_{n}$.

After this paper was completed, a detailed variational Monte-Carlo calculation
for the tetraneutron was done in Ref. \cite{Pie03}. It was shown that it does
not seem possible to change modern nuclear Hamiltonians to bind a tetraneutron
without destroying many other successful predictions of those Hamiltonians.
Otherwise, our understanding of nuclear forces would have to be significantly changed.

More elaborate QCD calculations based on lattice gauge theories\ cannot
presently assess this problem. There are some lattice calculations for the
H-dibarion state with not very reliable results (see Ref. \cite{WK02,Neg98}).
However, lattice claculations still cannot determine if, e.g., the deuteron is
bound. State of the art lattice numerical calculations are aiming at
$\mathcal{O}$(10) MeV accuracy in binding energy, not $\mathcal{O}$(1) MeV or
less that is relevant for nuclear physics. Also, how the calculations depend
on quark masses is an interesting but very difficult question.

\bigskip

We acknowledge useful discussions \ with M. Creutz, S. Pieper, S. Ohta and K.
Orginos. This work was supported by the National Science Foundation under
Grants No. PHY-0110253, PHY-9875122, PHY-007091, PHY-0070818 and PHY-0244453.

\end{document}